\documentclass[useAMS,usenatbib,usegraphicx]{mn2e}

\def\nodata{...}
\def\p1{\phantom{1}}

\def\simless{\mathbin{\lower 3pt\hbox
     {$\rlap{\raise 5pt\hbox{$\char'074$}}\mathchar"7218$}}}   
\def\simmore{\mathbin{\lower 3pt\hbox
     {$\rlap{\raise 5pt\hbox{$\char'076$}}\mathchar"7218$}}}   
\def\hide#1{}
\newcommand{\msun}{\ensuremath{{\rm M}_\odot}}

\title[Time lags in 4U 1608--52 and 4U 1636--53]{Time lags of the kilohertz quasi-periodic oscillations in the low-mass X-ray binaries 4U 1608--52 and 4U 1636--53}
\author[M. G. B. de Avellar, M. M\'endez, A. Sanna \& J. E. Horvath]
{Marcio G. B. de Avellar$^{1}$\thanks{marcavel@astro.iag.usp.br}, 
Mariano M\'endez$^{2}$, 
Andrea Sanna$^{2}$, 
\newauthor
Jorge E. Horvath$^{1}$\\
$^{1}$Instituto de Astronomia, Geof\'{i}sica e de Ci\^encias Atmosf\'ericas, Universidade de S\~ao Paulo, \\
Rua do Mat\~ao 1226, 05508-090, S\~ao Paulo, Brazil\\
$^{2}$Kapteyn Astronomical Institute, University of Groningen, P.O. Box 800, 9700 AV Groningen,\\
The Netherlands}
\begin{document}

\date{
}

\pagerange{\pageref{firstpage}--\pageref{lastpage}} \pubyear{2013}

\maketitle

\label{firstpage}

\begin{abstract}
We studied the energy and frequency dependence of the Fourier time lags and intrinsic coherence of the kilohertz quasi-periodic oscillations (kHz QPO) in the neutron-star low-mass X-ray binaries 4U 1608--52 and 4U 1636--53, using a large dataset obtained with the {\em Rossi X-ray Timing Explorer}. 
We confirmed that, in both sources, the time lags of the lower kHz QPO are soft (soft photons lag the hard ones) ranging from $\sim 10\mu$s to $100\mu$s, with the soft lags increasing with energy. Furthermore, we found that:
(i) In both sources, the time lags of the upper kHz QPO are independent of energy, and inconsistent with the soft lags of the lower kHz QPOs. (ii) In 4U 1636--53, for the lower kHz QPO the $4-12$ keV photons lag the $12-20$ keV ones by $\sim 25\mu$s in the QPO frequency range $500 - 850$ Hz, with the soft lags decreasing to $\sim 15\mu$s when the QPO frequency increases further. In 4U 1608--52 the soft lags of the lower kHz QPO remain constant at $40\mu$s up to $800$ Hz, the highest frequency reached by this QPO in our data. (iii) In 4U 1636--53, the time lags of the upper kHz QPO are hard (the $12-20$ keV photons lag the $4-12$ keV ones),  at $11 \pm 3\mu$s, independent of QPO frequency (from $500$ Hz to $1200$ Hz). We found consistent results for the time lags of the upper kHz QPO in 4U 1608--52, albeit with larger error bars than in 4U 1636--53 over the narrower frequency interval covered by the upper kHz QPO in our data ($880 - 1040$ Hz). (iv) In both sources the intrinsic coherence ($4 - 12$ keV vs. $12- 20$ keV) of the lower kHz QPO remains constant at $0.6$ between $5$ and $12$ keV, and drops to zero above that energy. The intrinsic coherence of the upper kHz QPO is consistent with being zero across the full energy range. (v) The intrinsic coherence of the lower kHz QPO increases from $\sim0-0.4$ at $\sim 600$ Hz to $1$ and $0.6$ at $800$ Hz in 4U 1636--53 and 4U 1608--52, respectively. In 4U 1636--53 the intrinsic coherence decreases to $0.5$ at $920$ Hz, while in 4U 1608--52 we do not have data above $800$ Hz. (vi) In both sources the intrinsic coherence of the upper kHz QPO is consistent with zero over the full frequency range of the QPO, except in 4U 1636--53 between $700$ Hz and $900$ Hz where the intrinsic coherence marginally increases, reaching a value $0.13$ at $780$ Hz. We discuss our results in the context of scenarios in which the soft lags are either due to reflection off the accretion disc or up-/down-scattering in a hot medium close to the neutron star. We finally explore the connection between, on one hand the time lags and the intrinsic coherence of the kHz QPOs, and on the other the QPOs' amplitude and quality factor in these two sources.
\end{abstract}


\clearpage

\section{Introduction}
\label{intro}

The kilohertz quasi-periodic oscillations (kHz QPOs) are the fastest variability so far observed from any astrophysical object. These QPOs are relatively narrow peaks in the power density spectra of neutron-star (NS) low-mass X-ray binaries (LMXBs). These peaks often appear in pairs at frequencies $\nu_1$ and $\nu_2 > \nu_1$ that change with time, ranging from $\sim 400$ Hz up to $\sim 1300$ Hz \citep[see][and references therein; we will refer to the kHz QPO at $\nu_1$ and $\nu_2$ as the lower and the upper kHz QPO, respectively]{vanderklis01}. Since the Keplerian frequency close to the innermost stable circular orbit (ISCO) around a $\sim 15$ km and $\sim 1.4$ \msun\ NS is $\sim 1300$ Hz, several models have been proposed in which the kHz QPOs, and especially the one at $\nu_2$, reflect the Keplerian frequency of matter orbiting at the inner edge of an accretion disc, very close to the NS  surface \citep[e.g.][]{miller01,stella01}. 

Most of the observational and theoretical studies in the past 15 years concentrated on the frequencies \citep[e.g.][]{psaltis01,straaten02,diego01, sanna02} and, to a lesser degree, the amplitude and the coherence of the kHz QPOs \citep[][see \citealt{vanderklis01} for a review]{mendez06, disalvo01, barret02, barret03, barret04, mendez04, sanna01}. Almost no work has been done on another aspect of the kHz QPO signal: The energy-dependent time (or phase) lags and the Fourier coherence \citep{nowak01} of the kHz QPOs.

The time (phase) lags and Fourier coherence are Fourier-frequency-dependent measures of, respectively, the time (phase) delay and the degree of linear correlation between two concurrent and correlated signals, in this case light curves of the same source, in two different energy bands. Perfect (unit) intrinsic (noise subtracted) coherence entails that one can predict one of the signals from the measurements of the other \citep[see][for a detailed explanation of the time lags, coherence, and intrinsic coherence, and how to calculate them]{nowak01}. There are only two papers in the literature that describe observational results of the time lags of the kHz QPOs: Using data from the {\em Rossi X-ray Timing Explorer (RXTE)}, \citet[][see also \citealt{vaughan02} for an erratum]{vaughan01} studied the time lags of the kHz QPOs in the NS LMXBs 4U 1608--52, 4U 1636--53 and 4U 0614+09, while \citet{kaaret01} did a similar work for the NS LMXB 4U 1636--53. These authors found that in all these sources, within a narrow frequency interval centred around the centroid frequency of the QPOs, the low-energy (soft) photons lag the high-energy (hard) photons by $\sim 5 - 100 \mu$s, with the magnitude of the lags increasing as the energy difference between soft and hard photons increases. Those two works were done at the beginning of the {\em RXTE} mission, when only a handful of observations for each of these sources were available; furthermore, the analysis had to be done using a coarse energy resolution to alleviate the limited signal-to-noise ratio of the data, and in some cases the lags could not be measured accurately or significantly. More importantly, at the time it was not possible to study the potential dependence of the time lags upon the frequency of the kHz QPO. Surprisingly, no similar observational study has been done since then, despite the fact that the {\em RXTE} archive contains enough data that could be used to explore this.

Here we extend the initial work by \citet{vaughan01,vaughan02} and \citet{kaaret01}, using a large set of {\em RXTE} observations to analyse the energy-dependent time lags and coherence of the kHz QPOs in two of the three NS LMXBs mentioned above: 4U 1608--52 and 4U 1636--53. 4U 1608--52 is a transient NS LMXB showing outbursts with recurrence times that vary from 80 days to a few years \citep{lochner01}. KHz QPOs have been detected in this source with frequencies ranging from $\sim 500$ Hz to $\sim 1050$ Hz \citep{berger01,mendez01,straaten01}. \citet{vaughan01,vaughan02} measured time lags in one of the kHz QPO \citep[later on identified as the lower kHz QPO; see][]{mendez01} in this source, going from $\sim -10\mu$s at $\sim 7$ keV to $\sim -60\mu$s at $\sim 23$ keV. (All time lags were calculated with respect to photons in the $4-6$ keV energy band. Because of this convention, a negative lag means that the hard photons lead the soft ones. These are usually called soft lags, as opposed to hard lags when the hard photons lag the soft ones.) 4U 1636--53 is a persistent NS LMXB that shows kHz QPOs with very similar properties (frequency range, amplitude and coherence) to the kHz QPOs in 4U 1608--52 \citep{mendez07,disalvo01,barret04}. \citet{kaaret01} measured soft lags in the lower kHz QPO in this source ranging from $\sim -5\mu$s at $\sim 7$ keV to $\sim -40\mu$s at $\sim 23$ keV, relative to the $3.8-6.4$ keV band. The only measurement of a time lag of the upper kHz QPO is that by \citet{vaughan01} who found a time lag of $-26 \pm 23 \mu$s between the $6.5-67$ keV and $2-6.5$ keV photons for the upper kHz QPO in 4U 0614+091.

%
%
%
\begin{table}
\caption{Mean energy of the different energy bands used to calculate the time lags of the kHz QPOs in 4U 1608--52 and for 4U 1636--53. In the case of 4U 1608--52 we show separately the mean energy in the two observing modes (32M and 64M; see \S\ref{obs}).}
\centering
\begin{tabular}{|c|c|c|}
\hline
\multicolumn{2}{|c|}{4U 1608--52} & \multicolumn{1}{|c|}{4U 1636--53}\\
\hline
$\bar{E}$ [keV] (32M)& $\bar{E}$  [keV] (64M)& $\bar{E}$ \\
\hline
  3.75     &   3.75 &   4.2 \\
  5.56     &   5.42 &   6.0 \\
  7.96     &   7.34 &   8.0 \\
10.94     &   9.47 & 10.2 \\
14.33     & 12.29 & 12.7 \\
17.49     & 15.86 & 16.3 \\
 \nodata & 18.10 & 18.9 \\
\hline
\end{tabular}
\label{tab1}
\end{table}

%
%
%
\begin{table*}
\begin{minipage}{126mm}
\caption{Frequency intervals, mean frequency, and number of 16-s segments for the lower and the upper kHz QPO in 4U 1608--52 and 4U 1636--53. In the case of 4U 1608--52 we show separately the number of 16-s segments in the two observing modes (32M and 64M).}
\centering
\begin{tabular}{|c|c|c|c|cr}
\hline
\multicolumn{3}{|c|}{4U 1608--52} & \multicolumn{3}{|c|}{4U 1636--53}\\
\hline
\multicolumn{6}{c}{Lower kHz QPO}\\
\hline
$\nu$ range [Hz] & $\bar{\nu}$ [Hz] (32M $\mid$ 64M) & \#16-s segments (32M $\mid$ 64M) &  $\nu$ range [Hz] & $\bar{\nu}$ [Hz] & \#16-s segments \\
\hline
    $  540 -   610$ & 579 $\mid$ \nodata ~~    & 635 $\mid$ \nodata ~~ &  $  550 -   620$ &  596   &      728 \\
    $  610 -   690$ & 655 $\mid$ 677                      & 553 $\mid$ 238     &  $  620 -   670$ &  650   &     6258 \\
    $  690 -   770$ & 745 $\mid$ 735                      & 582 $\mid$ 399     &  $  670 -   715$ &  693   &     6507 \\
    $  770 -   823$ & 791 $\mid$ 783                      & 616 $\mid$ 309     &  $  715 -   750$ &  731   &     6315 \\
    \nodata            & \nodata & \nodata &  $  750 -   790$ &  768   &    6515 \\
    \nodata            & \nodata & \nodata &  $  790 -   820$ &  807   &    7920 \\
    \nodata            & \nodata & \nodata &  $  820 -   850$ &  836   &    9436 \\
    \nodata            & \nodata & \nodata &  $  850 -   880$ &  864   &    9548 \\
    \nodata            & \nodata & \nodata &  $  880 -   910$ &  895   &    8954 \\
    \nodata            & \nodata & \nodata &  $  910 -   975$ &  921   &    3710 \\
\hline
\multicolumn{6}{c}{Upper kHz QPO}\\
\hline
$\nu$ range [Hz]  & $\bar{\nu}$ [Hz] (32M $\mid$ 64M) & \#16-s segments (32M $\mid$ 64M) &  $\nu$ range [Hz] & $\bar{\nu}$ [Hz] & \#16-s segments \\
\hline
    $  \phantom{1}859 -   \phantom{1}910$ & \phantom{1}887  $\mid$ \phantom{1}\nodata~~    & 611 $\mid$ \nodata~~       &  $  \phantom{1}440 -   \phantom{1}540$ &  \phantom{1}492	  &	7052 \\
    $  \phantom{1}910 -   \phantom{1}993$ & \phantom{1}952  $\mid$ \phantom{1}977                     & 549 $\mid$ 176     &  $  \phantom{1}560 -   \phantom{1}650$ &  \phantom{1}605	  &	4120 \\
    $  \phantom{1}993 -  1040$ & 1035 $\mid$ 1028                    & 604 $\mid$ 353     &  $  \phantom{1}650 -   \phantom{1}750$ &  \phantom{1}702	  &	5840 \\
    $ 1040 -  1064$ & 1047 $\mid$ 1044                    & 622 $\mid$ 417     &  $  \phantom{1}750 -   \phantom{1}810$ &  \phantom{1}781	  &    10871 \\
    \nodata            & \nodata & \nodata &  $  \phantom{1}810 -   \phantom{1}870$ &  \phantom{1}840	  &    8368 \\
    \nodata            & \nodata & \nodata &  $  \phantom{1}870 -   \phantom{1}930$ &  \phantom{1}897	  &	8544 \\
    \nodata            & \nodata & \nodata &  $  \phantom{1}930 - 1025$ &  \phantom{1}979	  &  11685 \\
    \nodata            & \nodata & \nodata &  $1070 - 1165$ & 1117	  &    7577 \\
    \nodata            & \nodata & \nodata &  $1165 - 1250$ & 1211	  &  12595 \\
\hline
\end{tabular}
\label{tab2}
\end{minipage}
\end{table*}

The origin of the time lags in the power density spectra of black-hole \citep{miyamoto01, nowak01} and neutron-star \citep{ford01} LMXBs, and super-massive black holes in active galactic nuclei \citep{zoghbi01} is not fully understood. The mechanisms proposed so far involve Compton up-/down-scattering of photons produced in the accretion disc (and in the case of neutron stars, on the neutron-star surface or boundary layer) in a corona of hot electrons that surrounds the system \citep{sunyaev01,payne01,lee01,lee02,falanga01} or, in the case of soft lags in active galactic nuclei, reflection of hard photons from the corona off the inner parts of the accretion disc \citep{zoghbi01,zoghbi02}. If these mechanisms are applicable to neutron-star systems, the time lags of the kHz QPOs provide a powerful tool to study the physics of the accretion flow in the vicinity of the neutron star, since these lags encode information about the size and the geometry of the scattering medium or the reflector. From the implied light-crossing time, the observed time lags must be produced within a very small region close to the neutron star surface. Here we combine {\em RXTE} data of each of these two sources separately, using a technique (see \S\ref{obs}) that allows us to increase the signal-to-noise ratio of the time lags and Fourier coherence, and at the same time enables us to study, for the first time, the dependence of the time lags upon the frequency of the kHz QPO. We present our results in \S\ref{res}, and discuss them in the context of existing models in \S\ref{dis}.

\section{Observations and Data Analysis}
\label{obs}

For this paper we used data of 4U 1608--52 and 4U 1636--53 taken with the Proportional Counter Array \citep[PCA;][]{jahoda01} on board {\em RXTE} \citep[][]{bradt01}. For each observation (see below) we calculated the complex Fourier transform in a number of energy bands. In all cases we calculated Fast Fourier Transforms up to a Nyquist frequency of 2048 Hz over segments of 16 s, yielding a minimum frequency and a frequency resolution of $1/16$ Hz. For each of these 16-s segments we stored the real and imaginary part of the Fourier transform as a function of frequency. We did this for all available observations.

For the analysis of 4U 1608--52 we used data from the 1998 outburst of the source. These are the same observations used by \citet{mendez01}. From those data we only used the observations in which the source showed kHz QPOs (observations marked with circles in Fig. 1 of \citealt{mendez01}). The data were taken using two different observing modes in which the full energy band of the PCA (nominally from 2 to 60 keV) was either divided into 32 or 64 energy channels (32M and 64M; see Table \ref{tab1}). Due to the ageing of the PCA instrument, the channel-to-energy conversion of the PCA changed over time. Those changes, however, took place over relatively long time scales (of the order of years; see for instance \citealt{straaten01}), except for a few occasions in which the mission control changed the gain of the PCA instrument manually\footnote{See the channel-to-energy conversion table for the PCA at http://heasarc.gsfc.nasa.gov/docs/xte/r-c\_table.html}. Because the observations of 4U 1608--52 that we used here span a time interval of less than 30 days, and during that period no manual gain changes were applied to the PCA, we could use always the same channel scheme to define the energy bands in our analysis (Table \ref{tab1}). 

In the case of 4U 1636--53 we used all the data available in the {\em RXTE} archive for this source up to May 2011; these are the same data used by \citet{sanna02}. Here we only selected the 528 observations of 4U 1636--53 in which \citet{sanna02} found at least one of the kHz QPOs in the power density spectrum of this source. The {\em RXTE} archive contains observations of 4U 1636--53 in different modes that sampled the full PCA band in different ways and with different number of channels. These observations were carried out over a period of about 15 years, during which the gain of the instrument changed significantly (both due to manual changes as well as to the ageing of the instrument itself). We therefore had to adjust our channel selections for different groups of observations, depending on the epoch of the observation, to always select more or less the same energy bands for our analysis. We give the average energy in each of the bands that we used for our analysis of both sources in Table \ref{tab1}.

We used the results of \citet{mendez01} and \citet{sanna02}, who measured the frequency of the lower kHz QPO over different time segments, to assign a lower QPO frequency to each 16-s segment for which we calculated the Fourier transform. When those authors measured the lower kHz QPO frequency over a segment longer than 16 s, we assigned the same lower QPO frequency to all our 16-s segments that were included within their time segment. We further assigned the same lower QPO frequency to the same 16-s time segment for all the Fourier transforms calculated in the different energy bands. 

%
%
%
\begin{figure}
\centering
\includegraphics[width=0.4\textwidth,angle=0]{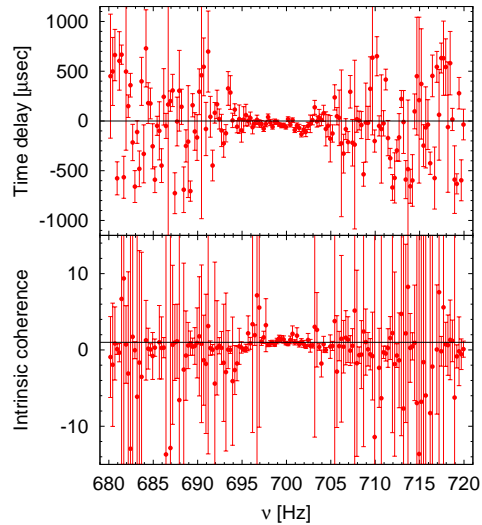}
\caption{Upper panel: Time lags as a function of Fourier frequency in 4U 1608--52, showing only the frequency range around 700 Hz. We calculated the time lags for photons with energies around 7.96 keV with respect to photons with energies around 3.75 keV (see Table \ref{tab1} for the definition of the energy bands), when the frequency of the lower kHz QPO was between 690 Hz and 770 Hz (see Table \ref{tab2} for the definition of the frequency intervals). To calculate the time lags we first shifted the frequency scale of the 16-s Fourier cross spectra in the selected frequency interval such that the lower kHz QPO was always centred at 700 Hz, and then calculated the average cross spectrum of all 16-s segments that belonged to that selection. The horizontal line is at a zero time lag. Lower panel: Intrinsic Fourier-coherence as a function of Fourier frequency in 4U 1608--52, calculated for the same energy bands and using the same selection of frequencies of the lower kHz QPO as in the upper panel. As in the upper panel, for this plot we also applied the shift-and-add technique to align the lower kHz QPO to a reference frequency of 700 Hz. The horizontal line shows the perfect intrinsic coherence of 1.}
\label{fig1}
\end{figure}

We also assigned an upper QPO frequency to each 16-s segment. We did this as follows: For the case of 4U 1608--52 we used the plot in Figure 3 of \citet{mendez01} that shows $\Delta \nu = \nu_2 - \nu_1$ vs. $\nu_1$ for the observations that we used in this paper. For each point in that Figure we added $\nu_1 + \Delta \nu$ to obtain $\nu_2$ for the corresponding value of $\nu_1$. We then assigned that upper kHz QPO frequency to all 16-s segments that were used by \citet{mendez01} to get those measurements of $\nu_1$  and $\Delta \nu$. For the case of 4U 1636--53 we used the measurements of \citet{sanna02} of the upper kHz QPO per individual observation (typically $3000 - 5000$ s exposure time), and we assigned that upper QPO frequency to all 16-s segments that belonged to that observation. As in the case of the lower kHz QPO, for each source we assigned the same upper QPO frequency to the same 16-s time segment in all the Fourier transforms calculated in the different energy bands.

In summary, we assigned a lower and an upper kHz QPO frequency to each 16-s segment for which we calculated the Fourier transform, the same lower and upper kHz QPO frequency per 16-s segment for each of the energy bands shown in Table \ref{tab1}. 

In our observations of 4U 1608--52 the lower and upper kHz QPO covered the frequency range of $\sim 540-820$ Hz and $\sim 860-1060$ Hz, respectively. In the case of 4U 1636--53, the lower and the upper kHz QPO covered the frequency range of $\sim 550-980$ Hz and $\sim 440-1250$ Hz, respectively. We therefore defined a number of frequency intervals for the lower and the upper kHz QPO for each source; we show the selected frequency intervals in Table \ref{tab2}. We chose the frequency intervals such that they spanned a sufficiently small frequency range to minimise the effect of possible frequency-dependent changes of the time lags of the kHz QPO, but large enough to have a sufficiently large number (balanced among the frequency intervals) of 16-s segments in each frequency interval to reduce the statistical uncertainties. 

%
%
%
\begin{figure*}
\centering
{
    \includegraphics[width=0.45 \textwidth, angle=0]{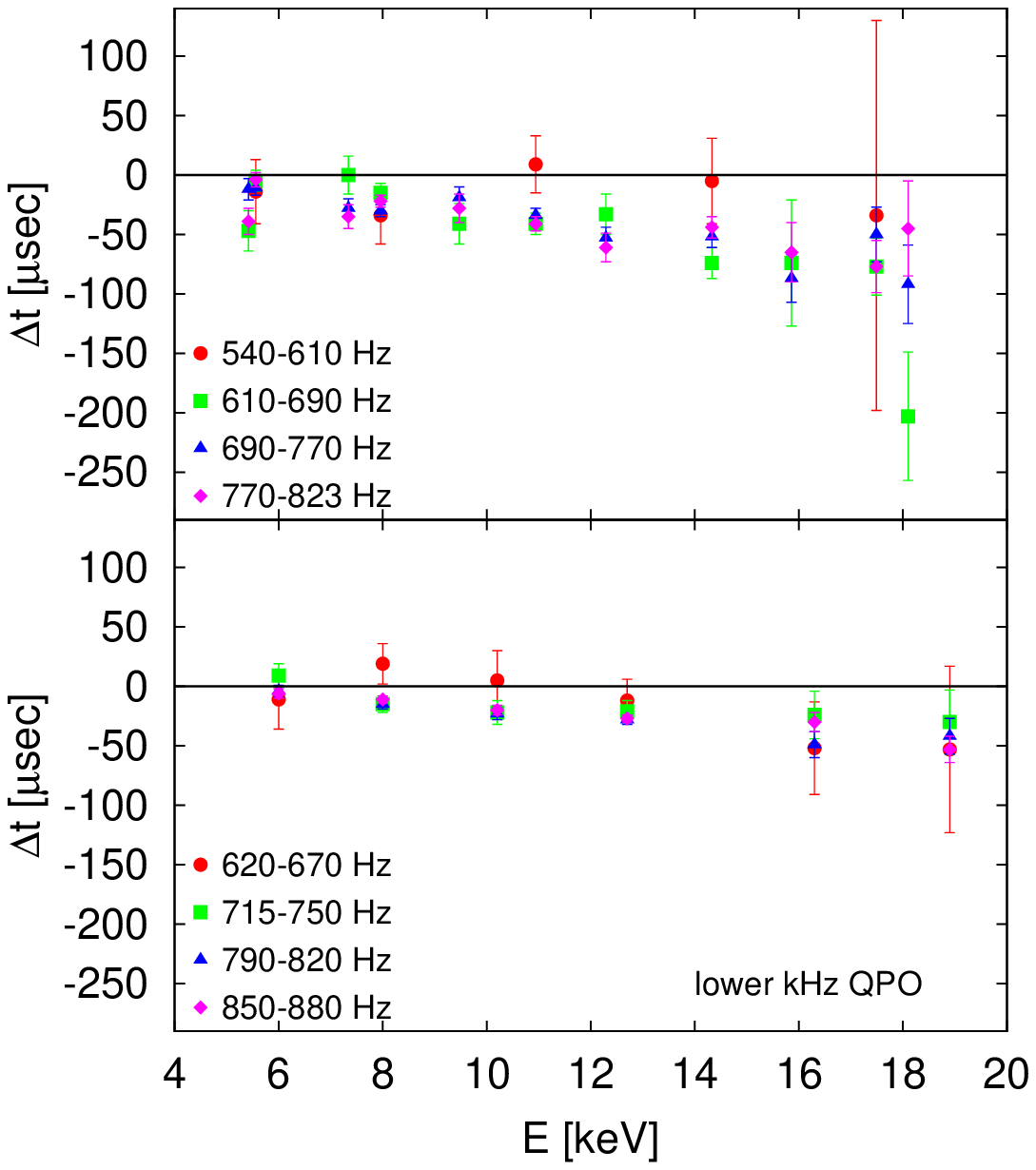}
}
{
    \includegraphics[width=0.45 \textwidth,angle=0]{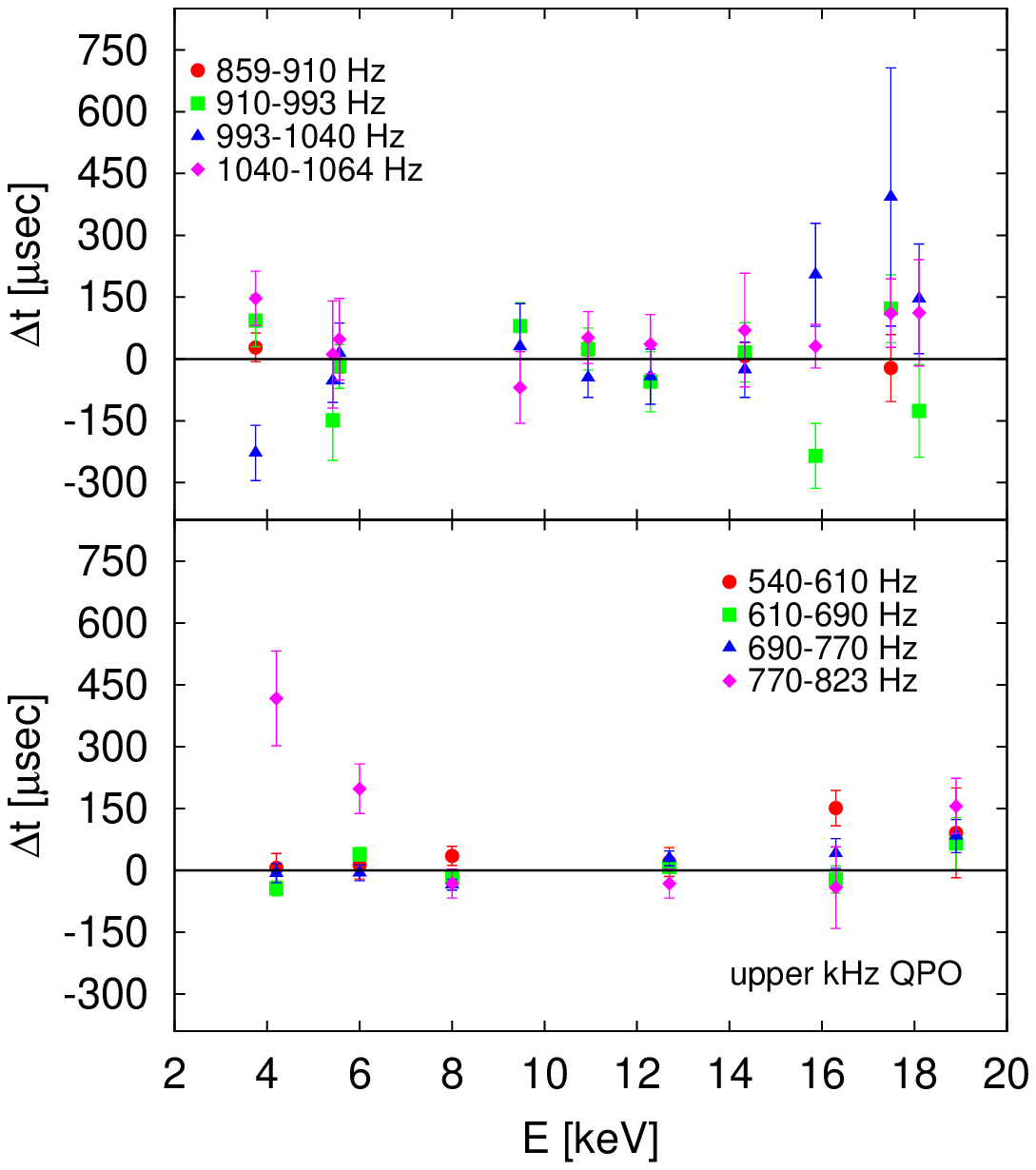}
}
\caption{Time lags vs energy of the lower (left column) and the upper (right column) kHz QPO in 4U 1608--52 (upper row) and 4U 1636--53 (lower row). We calculated the time lags between photons with energies given by the value on the $x$ axis relative to photons with energies around 3.75 keV and $\sim 7.5$ keV for the lower and the upper kHz QPO in 4U 1608--52, respectively, and 4.2 keV and 10.2 keV for the lower and the upper kHz QPO in 4U 1636--53, respectively. The different colours and symbols show the time lags when the frequency of the QPO was within the intervals indicated in the Figures. The horizontal line in all panels is at a zero time lag.}
\label{fig2} 
\end{figure*}

For each source, and for each kHz QPO separately, we applied the shift-and-add technique \citep{mendez01} to check the significance of the QPO in the power density spectra for different selections of the data. In both sources, both kHz QPOs were highly significant when we produced a single power density spectrum from all observations with kHz QPOs combined, shifting and adding the power density spectra of the individual time segments such that the frequency of either the lower or the upper kHz QPO was always the same. Similarly, both kHz QPOs were highly significant in both sources when we created shifted-and-added power density spectra for each frequency interval defined in Table \ref{tab2}, combining the photons of all energy bands, as well as when we created shifted-and-added power density spectra for each energy band defined in Table \ref{tab1} combining the data of all the frequency intervals in Table \ref{tab2}. We did not always detect the kHz QPOs significantly when we created power density spectra dividing the data both in energy and frequency. Given that we have $\sim 20$ times more data for 4U 1636--53 than for 4U 1608--52, this problem occurred more often in the latter than in the former source. In general, we detected the kHz QPOs more significantly when we selected energies between $\sim 5$ keV and $\sim 16$ keV and the frequency of the lower and the upper kHz QPO were, respectively, in the range $\sim 600 - 850$ Hz and $\sim 500 - 1000$ Hz \citep[see][]{berger01, barret04}. For those frequency and energy selections for which we did not detect either of the kHz QPOs significantly, we measured the time lags and intrinsic coherence of the QPOs (see below) within a frequency range centred around the QPO that we detected in other cuts of the data.  By doing this we implicitly assumed that the QPO frequency is independent of energy \citep[see, e.g.][]{berger01}.

We then applied the same idea behind the shift-and-add technique introduced by \citet{mendez01} to study the kHz QPOs in Fourier power density spectra, but this time to study the complex Fourier spectra, and in particular the time lags and the intrinsic coherence of the kHz QPOs in these two sources. To calculate the time lags and the intrinsic coherence of the kHz QPOs we followed the same procedure described by \citet{vaughan01} and \citet{kaaret01}, but in this case we first aligned the frequency axes of the Fourier transforms of each 16-s segment before we calculated the average cross spectra. To do this, for each frequency interval and for each source separately, we first shifted the frequency scale of the Fourier spectra such that in each frequency selection first the frequency of the lower and then that of the upper kHz QPO was always the same. For each source we also calculated the average frequency of the QPO for each frequency interval (Table \ref{tab2}), separately for the lower and the upper kHz QPO. We then selected one energy band from the ones given in Table \ref{tab1} as the reference band, and we calculated average cross spectra between each of the other energy bands in Table \ref{tab1} and this reference band. It is custom to choose the lowest energy band available as the reference band, but since the time lag is a relative quantity, one is free to choose any band for that. The accuracy with which one can measure the time lags depends upon the intensity and the strength of the variability of the signal \citep{vaughan03}, therefore it is convenient to choose a reference band in which both the variability and the source intensity are high. While it is easier to choose always the same reference band, in our case we were limited by the fact that the two sources in this paper were observed using different observing modes, and the instrument calibration changed with time (see above); this prevented us from defining exactly the same energy bands for the analysis of both sources. We therefore chose the reference band as follows, unless otherwise indicated: For the lower kHz QPO in 4U 1608--52 all time lags are calculated relative to the band centred at 3.75 keV, both for data observed in the 32M and 64M mode. For the upper kHz QPO in 4U 1608--52 we used the band centred at 7.96 keV for the 32M data and the band centred at 7.34 keV for the 64M data. To simplify the description, we will refer to the bands just mentioned as the band centred at $\sim 7.5$ keV. For the lower and the upper kHz QPO in 4U 1636--53 we used the bands centred at 4.2 keV and 10.2 keV, respectively. 

Finally, to calculate the time lags and intrinsic coherence of the lower kHz QPO we averaged the energy- and frequency-dependent time lags and intrinsic coherence over a frequency interval of 16 Hz centred around the centroid frequency of the lower kHz QPO; this range corresponds to about twice the full-width at half-maximum, FWHM, of the lower kHz QPO in the combined shifted-and-added power density spectrum of all observations with a lower kHz QPO. Similarly, to calculate the time lags and intrinsic coherence of the upper kHz QPOs we averaged the energy- and frequency-dependent time lags and intrinsic coherence over a frequency interval of 140 Hz centred around the centroid frequency of the upper kHz QPO; this range corresponds to $\sim 1.2$ times the FWHM of the upper kHz QPO in the combined shifted-and-added power density spectrum of all observations with an upper kHz QPO. Formally these are the time lags and intrinsic coherence of the light curves in the two energy bands used in the calculation, over the time scale (the inverse of the Fourier frequency) of either the lower or the upper kHz QPO. For simplicity, in the rest of the paper we call these the time lags and intrinsic coherence of either the lower or the upper kHz QPO.

\section{Results}
\label{res}

In the upper panel of Figure \ref{fig1} we plot the time lags as a function of Fourier frequency in 4U 1608--52 for the photons around 7.96 keV with respect to the photons around 3.75 keV, for the lower kHz QPO in the frequency interval between 690 Hz and 770 Hz. 
The plot only shows the frequency range around 700 Hz, and it reveals that the time lags have large statistical fluctuations (see the large error bars; throughout the paper we always quote 1-$\sigma$ errors), except in the interval between $\sim 695$ Hz and $\sim 705$ Hz, around the frequency that we used as reference to shift the lower kHz QPOs. The region where the time lags have small errors spans about twice the FWHM of the lower kHz QPO in the shifted-and-added power density spectrum of this selection. Notice that the time lags around the frequency of the QPO are negative, following the convention mentioned in \S\ref{intro} that negative values indicate soft lags.

%
%
%
\begin{figure}
\centering 
\includegraphics[width=0.4 \textwidth, angle=0]{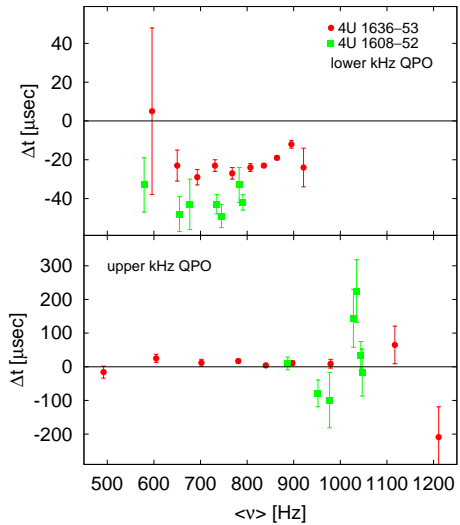}
\caption{Time lags vs. QPO frequency for photons with energy in the 12--20 keV band with respect to photons with energy in the 4--12 keV band for the lower (upper panel) and the upper kHz QPO (lower panel) of 4U 1608--52 (green squares) and 4U 1636--53  (red circles). The horizontal line in both panels is at a zero time lag.}
\label{fig3}
\end{figure}

%
%
%
\begin{figure}
\centering
\includegraphics[width=0.4 \textwidth,angle=0]{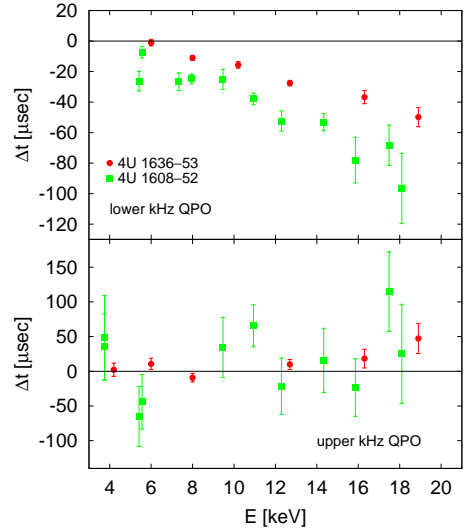}
\caption{Time lags vs. energy for the lower and the upper kHz QPOs in 4U 1608--52 (green squares) and 4U 1636--53 (red circles), for all frequency intervals combined. Upper panel: Time lags of the lower kHz QPO for photons with energies given by the value on the $x$ axis with respect to photons with energies around 3.75 keV and 4.2 keV for 4U 1608--52 and 4U 1636--53, respectively.  Lower panel: Time lags of the upper kHz QPO for photons with energies given by the value on the $x$ axis with respect to photons with energies around $\sim 7.5$ keV and 10.2 keV for 4U 1608--52 and 4U 1636--53, respectively. The horizontal line in both panels is at a zero time lag.} 
\label{fig4} 
\end{figure}

In the lower panel of Figure \ref{fig1} we plot the intrinsic Fourier coherence as a function of Fourier frequency for 4U 1608--52, calculated using the same two energy bands and for the same frequency interval of the lower kHz QPO, 690 Hz and 770 Hz, as in the upper panel of this Figure. 
As in the case of the upper panel of this Figure, it is apparent that the intrinsic coherence is not well defined (it shows large fluctuations and very large error bars), except in a narrow range around the frequency that we used as reference to shift the lower kHz QPO. The coherence is well defined (it has small error bars) at around a value of 1 within about twice the FWHM of the kHz QPO.

%
%
%
\begin{figure*}
\centering
{
        \includegraphics[width=0.7 \textwidth, angle=0]{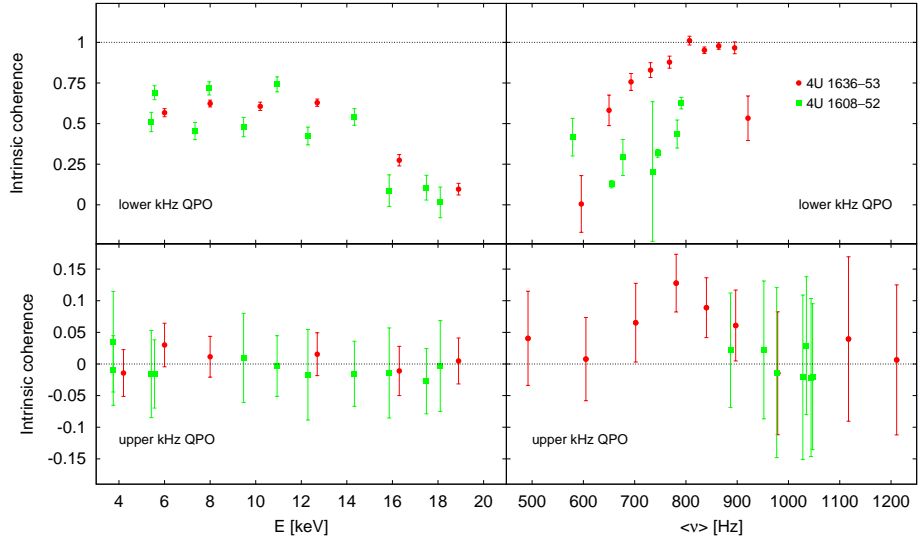}
}
\caption{Intrinsic Fourier coherence vs. energy (left panels) and vs. frequency (right panels) of the lower and the upper kHz QPOs (upper and lower panels, respectively) in 4U 1608--52 (green squares) and 4U 1636--53 (red circles). For the plot in the left panels we calculated the intrinsic coherence of the QPOs combining all frequency intervals, for photons with energies given by the value on the $x$ axis relative to photons with energies around 3.75 keV and $\sim 7.5$ keV for the lower and the upper kHz QPO in 4U 1608--52, respectively, and 4.2 keV and 10.2 keV for the lower and the upper kHz QPO in 4U 1636--53, respectively. For the plot in the right panels we calculated the intrinsic coherence of photons in the $4 - 12$ keV band relative to photons in the $12 - 20$ keV band. The horizontal lines indicate the intrinsic coherence values of 1 and 0 in the upper and the lower panels, respectively.}
\label{fig5} 
\end{figure*}

In Figure \ref{fig2} we plot the time lags as a function of energy for the lower (left panel) and the upper kHz QPO (right panel), both for 4U 1608--52 (upper panels) and 4U 1636--53 (lower panels). The points with different colours and symbols correspond to the time lags as a function of energy when the lower or the upper kHz QPO covered the frequency intervals indicated in the legend of the Figure. In all cases we calculated the time lags of the photons with average energy indicated in the $x$ axis with respect to photons in the reference band given in the previous section. For the lower kHz QPO in 4U 1608--52 (upper left panel), the soft photons lag the hard photons (which, according to the convention we use in this paper, yields negative lags in the plot) by $\sim 10 - 50\mu$s at $\sim 6$ keV, up to $\sim 100 - 200\mu$s at $\sim 18$ keV \citep[see][]{vaughan01,vaughan02}. For the lower kHz QPO in 4U 1636--53 (lower left panel) the soft photons lag the hard ones by $\sim 10\mu$s at $\sim 6$ keV up to $\sim 50\mu$s at $\sim 19$ keV \citep[see][]{kaaret01}. In both sources the time lags of the upper kHz QPO (right panels) have large errors, and are consistent both with zero lag and with the same trend with energy seen for the lower kHz QPOs. From this Figure it is also apparent that in both sources, and for each QPO separately, at each energy the time lags are consistent with being more or less the same in all frequency intervals, although the errors bars are relative large. 

In the two panels of Figure \ref{fig3} we plot the time lags of the lower and the upper kHz QPO in 4U 1608--52 and 4U 1636--53 for photons with energies $12-20$ keV with respect to the photons with energies $4-12$ keV as a function of the average QPO frequency in each frequency interval. The upper panel shows that the time lags of the lower kHz QPO are soft (the $12-20$ keV photons lead the $4-12$ keV photons) in both sources. In the case of 4U 1636--53 the time lags of the lower kHz QPO are consistent with being constant at $\sim -25\mu$s from $\sim 500$ Hz up to $\sim 850$ Hz, and then the magnitude of the lags decreases slightly, but significantly, to $\sim -10\mu$s as the frequency of the QPO increases further. The magnitude of the time lag of the last point, at $\sim 920$ Hz, appears to increase again but, given the error of that measurement, the value is also consistent with the overall trend of the previous points. A fit with a constant to the time lags over the full frequency range yields an average time delay of $-21.0 \pm 0.6\mu$s, but the fit is not good, yielding $\chi^2 = 39.5$ for 9 degrees of freedom (dof). The F-test probability for a fit with a linear function compared to a fit with a constant is 0.012. There is no significant improvement either if we fit with a parabola which, compared to a linear function, also yields an F-test probability of 0.012. A fit to the data with a parabola is marginally better than a fit with a constant, F-test probability of 0.002. The time lags of the lower kHz QPO in 4U 1608--52 are consistent with being constant, with an average time delay of $-42.9 \pm 0.3\mu$s  ($\chi^2 = 3.1$ for 6 dof), albeit with larger error bars than in the case of 4U 1636--53, but since the measurements do not extend beyond $\sim 800$ Hz, we cannot discard a similar trend as the one observed in 4U 1636--53 at the high end of the frequency range. 

The lower panel of Figure \ref{fig3} shows that in 4U 1636--53 the time lags of the upper kHz QPO are positive (the $12-20$ keV photons lag the $4-12$ keV photons) and independent of frequency. The average time lag of the upper kHz QPO in 4U 1636--53 over the whole frequency range is $11 \pm 3\mu$s ($\chi^2 = 12.9$ for 8 dof), almost $4$-$\sigma$ different from zero. The average time lag of the upper kHz QPO in 4U 1608--52 over the full frequency range is $5 \pm 15\mu$s ($\chi^2 = 14.9$ for 6 dof), both consistent with the average time lag in 4U 1636--53 (over a narrower frequency range, $890 - 1040$ Hz), and with zero lags. However, in both sources the average time lag of the upper kHz QPO is significantly different than the average time lag of the lower kHz QPO, and in 4U 1636--53, for which the statistics are better, it is apparent that the time lags of the lower and the upper kHz QPOs have opposite signs. In other words, while for the lower kHz QPO the soft photons lag the hard ones, for the upper kHz QPO the hard photons lag the soft ones.

Since the time lags of the lower and the upper kHz QPO, both in 4U 1608--52 and 4U 1636--53, appear not to change much with frequency (except for the increase of the lags of the lower kHz QPO in 4U 1636--53 mentioned above; see Figures \ref{fig2} and \ref{fig3}), for each of the kHz QPOs and for each source, we calculated the energy-dependent time lags over the whole frequency range spanned by each kHz QPO (i.e., no frequency selection). The upper panel of Figure \ref{fig4} shows that the magnitude of the soft time lags of the lower kHz QPO increases more or less linearly with energy in both sources (the best-fitting linear relation yields $\chi^2 = 9.8$ for 9 dof for 4U 1608--52 and $\chi^2 = 2.7$ for 4 dof for 4U 1636--53), with the magnitude of the soft time lags increasing marginally faster in 4U 1608--52, $5.0 \pm 0.5\mu$s/keV, than in 4U 1636--53, $3.6 \pm 0.3\mu$s/keV. The lower panel of Figure \ref{fig4} shows that in both sources the time lags of the upper kHz QPO are consistent with being zero (the average time lags being $13 \pm 13\mu$s and $4 \pm 3\mu$s for 4U 1608--52 and 4U 1636--53, respectively), independent of energy. (A joint fit to the data of both sources with a linear function, compared to a constant fit, gives an F-test probability of 0.08.)

In the left panel of Figure \ref{fig5} we plot separately the intrinsic coherence of the lower and the upper kHz QPO in 4U 1608--52 and 4U 1636--53 using all the frequency intervals combined. For this Figure we calculated the intrinsic coherence of photons with average energy indicated in the $x$ axis with respect to photons with energies around 3.75 keV and 4.2 keV for the lower kHz QPO in 4U 1608--52 and 4U 1636--53, respectively, and with respect to photons with energies around $\sim 7.5$ keV and 10.2 keV for the upper kHz QPO in 4U 1608--52 and 4U 1636--53, respectively. In both sources the intrinsic coherence of the lower kHz QPO remains constant at $\sim 0.6$ (where an intrinsic coherence of 1 indicates perfect correlation between the two signals) up to $14$ keV (upper left panel), and it drops to 0 (indicating that the signals are completely uncorrelated) above that energy. For both sources the intrinsic coherence of the upper kHz QPO is consistent with being zero across the full energy range (lower left panel). 

Finally, in the right panel of Figure \ref{fig5} we plot separately the intrinsic coherence of the lower and the upper kHz QPO in 4U 1608--52 and 4U 1636--53 as a function of the frequency of either of the kHz QPO. For this Figure we calculated the intrinsic coherence of photons in the $4-12$ keV band relative to photons in the $12-20$ keV band. For both sources the intrinsic coherence of the lower kHz QPO (upper panel) increases from $\sim 0-0.4$ at $\sim 600$ Hz to $\sim 1$ and $\sim 0.6$ at $\sim 800$ Hz in 4U 1636--53 and 4U 1608--52, respectively. In 4U 1636--53 the intrinsic coherence drops significantly above $\sim 900$ Hz, going down to $\sim 0.5$ at $\sim 920$ Hz. In 4U 1608--52 we do not have data above $\sim 800$ Hz to assess whether, as in the case of 4U 1636--53, the coherence also drops above that frequency. In both sources the intrinsic coherence of the upper kHz QPO (lower panel) is consistent with being zero over the full frequency range of the upper kHz QPO, except for 4U 1636--53 over a narrow frequency range between $\sim 700$ Hz and $\sim 900$ Hz where the intrinsic coherence marginally increases reaching a value of $\sim 0.13$ at $\sim 780$ Hz and then it drops again to 0 at $\sim 950$ Hz. In the case of 4U 1608--52 we do not have data below $\sim 880$ Hz, and hence we cannot exclude a similar behaviour to that seen in 4U 1636--53.

\section{Discussion}
\label{dis}

%
%

We studied the energy and frequency dependence of the time lags and intrinsic Fourier coherence of the kilohertz quasi-periodic oscillations in the neutron-star low-mass X-ray binaries 4U 1608--52 and 4U 1636--53, using data from the {\em Rossi X-ray Timing Explorer}. For the first time we analysed the dependence of the time lags and intrinsic coherence of the kHz QPOs upon the QPO frequency. For both sources, the time lags of the lower kHz QPO are soft (the $4-12$ keV photons lag the $12-20$ keV ones), and of the order of a few tens of microseconds. In 4U 1636--53 the soft time lags of the lower kHz QPO remain more or less constant at $25\mu$s as the QPO frequency changes from 500 Hz to 850 Hz whereas, when the QPO frequency increases further, the soft lags decrease to $10\mu$s. In 4U 1608--52 the soft QPO photons lag the hard ones by $40\mu$s, independent of QPO frequency from 580 Hz up to 780 Hz, the highest frequency reached by this QPO in our data of this source. In both sources the time lags of the upper kHz QPO are inconsistent with the time lags of the lower kHz QPO. In 4U 1636--53 the hard photons lag the soft ones, so-called hard lags, by $11 \pm 3\mu$s, independent of QPO frequency from $500$ Hz to $1200$ Hz. We found consistent results for the upper kHz QPO in 4U 1608--52, albeit with larger error bars than in 4U 1636--53 over the narrower frequency interval covered by the upper kHz QPO in our data, from 880 Hz to 1040 Hz. The intrinsic coherence ($4 - 12$ keV vs. $12- 20$ keV photons) of the lower kHz QPO increases from $\sim 0-0.4$ at $\sim 600$ Hz to $1$ and $\sim 0.6$ at $\sim 800$ Hz in 4U 1636--53 and 4U 1608--52, respectively. In 4U 1636--53 the intrinsic coherence decreases significantly to $0.5$ at $920$ Hz, while in 4U 1608--52 we do not have data above $800$ Hz. On the contrary, in both sources the intrinsic coherence of the upper kHz QPO is consistent with zero over the full frequency range of the QPO, except in 4U 1636--53 between $\sim 700$ Hz and $\sim 900$ Hz where the intrinsic coherence marginally increases to $0.13$ at $780$ Hz and then drops again to 0 at $\sim 950$ Hz. Since we do not have data below $880$ Hz, we cannot exclude a similar behaviour in the case of 4U 1608--52. We further studied the dependence of the time lags and intrinsic Fourier coherence of the kHz QPOs upon energy. This is the first such study in the case of the time lags of the upper kHz QPO and the intrinsic coherence of both the lower and the upper kHz QPO. We found that in both sources the intrinsic coherence of the lower kHz QPO remains constant at $\sim 0.6$ between $5$ and $\sim 12$ keV, and drops to zero above that energy. The intrinsic coherence of the upper kHz QPO in both sources is consistent with being zero across the full energy range. We confirmed, with finer energy resolution than in previous analysis \citep{vaughan01,vaughan02,kaaret01} that in both sources the soft time lags of the lower kHz QPO range from $\sim 10$  to $1\sim 00\mu$s, with the soft lags increasing with energy. In both sources, the time lags of the upper kHz QPO are consistent with zero and independent of energy.

%
%

%
%

Regardless of whether the time lags of the kHz QPOs are soft or hard (negative or positive, respectively, according to the convention we used in this paper), light travel time arguments indicate that the size of the region over which these time lags are produced must be smaller than $\sim 3 - 30$ km, depending on the energies used to measure the lags (see e.g., Figures \ref{fig3} and \ref{fig4}). The small change of the time lags with QPO frequency imply that the geometry of the region that produces the lags does not change much either. For instance, the fact that the time lags of the upper kHz QPO in 4U 1636--53 are consistent with being constant at $11 \pm 3\mu$s while the frequency of the QPO changes from $\sim 490$ Hz to $\sim 1210$ Hz, implies that the size of the region, or the path length, over which the QPO photons are delayed can at most change by $\sim 1$ km. Notice that if the frequency of the upper kHz QPO is the Keplerian frequency at the inner edge of the accretion disc \citep{miller01, stella01}, for a neutron-star mass of $\sim 1.4 - 2.2$ \msun\ the inferred inner-disc radius would change by $12 - 14$ km¤.

%
%

\citet{zoghbi01} proposed that the $\sim 30$-s soft lags seen in the active galactic nucleus 1H0707--495 at Fourier frequencies higher than $\sim 5\times10^{-4}$ Hz are due to reflection of photons from the corona around the central black hole off the accretion disc. If this scenario is correct and holds also for accreting neutron-star systems, besides photons from the corona one would also have to consider relatively hard photons coming from the neutron-star surface or boundary layer \citep{sunyaev02, kluzniak01} reflecting off the accretion disc \citep{cackett01, dai01, sanna03}. On the other hand, all models of the kHz QPOs propose that the frequency of either the lower or the upper kHz QPO is equal to the Keplerian frequency at the inner edge of the accretion disc \citep[e.g.][]{miller01, stella01, osherovich01}. In all these models, changes of the QPO frequency reflect changes of the inner radius of the accretion disc, set by the interplay between mass accretion rate, radiation pressure and radiation drag \citep{miller01, sanna02}. A combination of these two ideas (soft lags due to reflection off the accretion disc and one of the kHz QPOs being Keplerian), appears to be consistent (at least qualitatively) with our measurements of the magnitude of the time lags of the lower kHz QPO and their dependence on QPO frequency: In 4U 1636--53 the magnitude of the soft time lags of the lower kHz QPO decreases by $\sim 15\mu$s as the frequency of the QPO increases from  $\sim 600$ Hz to $\sim 900$ Hz (see Figure \ref{fig3}); the relative distance between the region in the corona or the neutron-star surface or boundary layer where the hard QPO photons are produced and the region in the disc where the soft QPO photons are reflected, would then change by $\sim 5$ km or less. If we identify the lower kHz QPO as the Keplerian frequency at the inner edge of the disc \citep{osherovich01}, the frequency range spanned by the QPO implies a change of the radius of the inner edge of the disc of $\sim 5 - 6$ km for neutron-star masses of $\sim 1.4 - 2.2$ \msun. If instead we identify the upper kHz QPO as the Keplerian one \citep{miller01, stella01}, using the typical frequency separation of $\sim 300$ Hz between the two simultaneous kHz QPOs in this source \citep{wijnands01, mendez07}, the frequency range spanned by the lower kHz QPO implies a change of the inner disc radius of $\sim 4$ km for the same range of neutron-star masses. Not only the magnitude of the soft time lags of the lower kHz QPO and their variation match the typical size and range of inferred inner-disc radii (assuming a localised source of hard photons, e.g. the neutron-star surface), but also the magnitude of the soft time lags decreases as the QPO frequency increases (and the inferred inner-disc radius decreases), which would be the expected behaviour if the bulk of the reflection takes place close to the inner edge of the disc. It is also interesting to note that in the case of 1H0707--495 the lags become negative at $\sim 5 \times 10^{-4}$ Hz, which scales up to $\sim 1200$ Hz if the mass ratio of the black hole in 1H0707--495 \citep{zoghbi01} and the neutron star in 4U 1636--453 \citep{barret04} is $\sim 5 \times 10^6 \msun / 2 \msun$. 

%
%

The time lags of the upper kHz QPO in 4U 1636--53 are hard, almost 4-$\sigma$ different from zero, and significantly different from the time lags of the lower kHz QPO in this source. The measurements in 4U 1608--52 have larger errors, but they are consistent with those in 4U 1636--53. It appears difficult to explain simultaneously the hard time lags of the upper kHz QPO and the soft time lags of the lower kHz QPO as reflection off the accretion disc around the neutron star. On one hand, the photons reflected off the disc would have to be harder than the incident photons from the corona or the neutron-star surface or boundary layer in the case of the upper kHz QPO whereas, in the scenario sketched in the previous paragraph for the lower kHz QPO, the reflected photons were assumed to be softer than the incident ones. If both the soft and hard time lags of, respectively, the lower and the upper kHz QPO are due to reflection off the accretion disc, this would require, for instance, that for the lower kHz QPO the incident photons came from the corona or relatively hot areas of the neutron-star surface or boundary layer and reflected off relatively cold areas in the disc, whereas for the upper kHz QPO the incident photons came from relatively cold areas in the neutron-star surface or boundary layer and reflected off relatively hot areas in the disc. This scenario seems unlikely, given that the disc temperature decreases with distance from the neutron star and hence, from travel time arguments, as the inner disc radius changes the magnitude of the time lags of the upper kHz QPO would have to change more than those of the lower kHz QPO, contrary to what we observed.

%
%

The oscillation mechanism that produces the quasi-periodic variability likely takes place in the disc; it is relatively straightforward to find characteristic (dynamical) frequencies in the accretion disc that match the observed frequencies of the kHz QPOs \citep[e.g.][]{miller01, stella01, osherovich01, abramowicz01}, whereas it is more difficult to find a `clock' somewhere else in the accretion flow. However, the emission from the disc alone cannot explain the rms amplitudes of the kHz QPOs, since in some cases the modulated luminosity in the QPOs is $\sim 15$ per cent, whereas at the same time the emission of the disc is $\simless 10$ per cent of the total luminosity of the source. The steep increase of the rms amplitude with energy implies a large modulation of the emitted flux, up to $\sim 20$ per cent at $\sim 25-30$ keV \cite[see e.g.][]{berger01}, at energies where the contribution of the disc is negligible. Hence, while the oscillation mechanism probably takes place in the disc, the signal is likely modulated and amplified by one of the high-energy spectral components in these sources \citep{mendez04, sanna01}. The interpretation that the time lags of the kHz QPO are due to reflection off the accretion disc needs to address the fact that the rms spectrum of the kHz QPOs is hard \citep{berger01, mendez06}, and that the frequency-resolved energy spectrum of the kHz QPOs matches the spectrum of the boundary layer in these systems \citep{gilfanov01}. We note that in 4U 1608--52 the rms spectrum of the lower kHz QPO is steeper than that of the upper kHz QPO \citep{mendez06}; it is unclear whether this is a general feature of the kHz QPOs, and how the rms spectrum of the kHz QPO depends, for instance, upon the frequency of the kHz QPO itself. This may be a relevant piece of information given the opposite sign of the time lags of the lower and the upper kHz QPO in 4U 1636-53. At any rate, quantitative calculations would be needed to assess whether specific geometries, either of the emitter or the reflector, could explain both the time lags and the rms spectra of both kHz QPOs under the assumption that the time lags of both kHz QPOs are due to reflection off the accretion disc in accreting neutron-star systems.

%
%

Other proposed scenarios for producing time lags, be them soft or hard, involve Compton up- or down-scattering \citep{lee01, lee02, falanga01}. For instance, \citet{falanga01} proposed a model for the time lags of the pulsations of accreting millisecond pulsars in which hard photons emitted from the accreting column or from a hot corona around the neutron star are Compton down-scattered by cold plasma in the the accretion disc or the neutron-star surface. In this model, time lags of the order of a few to a few tens of microseconds imply that the region where the time lags are produced must be small (see below). In principle, the same model would apply to the case of the kHz QPOs, as long as the emission and scattering regions are the same as in the case of millisecond pulsars. This would be the case if the QPO signal comes from the corona or the neutron-star surface or boundary layer, and the down-scattering medium is the atmosphere of the disc. If we fit the energy-dependent time lags of the lower kHz QPO (Figure \ref{fig4}) with this model, we find number density of the down-scattering medium of $\sim 1 - 2 \times 10^{20}$ cm$^{-3}$ in both sources, which is about four orders of magnitude smaller than the expected mid-plane density of a standard $\alpha$-disc extending close to the surface of a $1.7$-\msun\ and 10-km neutron star with a spin period of $\sim 600$ Hz and a magnetic field of $1 - 2 \times 10^8$ G \citep[see, for instance,][]{frank01}. For an optical depth $\tau= 1 - 5$, the size of the down-scattering region would be $\sim 100 - 700$ m. The fits with this model are, however, not good, with $\chi^2$ values of 42 for 9 dof and 68 for 4 dof for 4U 1608--52 and 4U 1636--53, respectively. If we apply the same model to the time lags of the upper kHz QPO we find that the number density of the up-scattering medium is $\sim 6 \times 10^{18}$ cm$^{-3}$ in both sources, about one order of magnitude lower than in the case of the lower kHz QPO; correspondignly, the size of the up-scattering region in this case would be $\sim 3 - 16$ km for $\tau=1 - 5$. While the fits are better than in the case of the lower kHz QPO, they are still formally unacceptable, with $\chi^2$ values of 12 for 9 dof and 6 for 4 dof for 4U 1608--52 and 4U 1636--53, respectively. One way to justify the low densities and, at least in the case of the lower kHz QPOs, the small size of this region would be if the disc is truncated, for instance at the sonic radius or the radius of the ISCO. In that case the radial velocity of the infalling material would increase rapidly there, and the density of the material would decrease sharply. 

%
%

For the broad-band noise component, between 0.01 and 100 Hz, in the power density spectra of neutron-stars and black-hole X-ray binaries hard X-rays lag soft X-rays \citep[e.g.][]{miyamoto01, nowak01, ford01}. In those cases the time lags decrease more or less linearly with frequency (the phase lags, equal to the time lags multiplied by the Fourier frequency, are more or less constant as a function of Fourier frequency). These broad-band frequency-independent hard phase (time) lags have been interpreted in terms of a Compton up-scattering medium with a radial density gradient \citep{kazanas01, hua01}. The idea behind the Compton up-scattering model is that low-energy photons gain energy in each scattering event. More scatterings imply a higher energy of the emerging photons, but also a longer path length, therefore the higher energy photons emerge later than the lower energy photons produced simultaneously. It is intriguing that the time lags of the broad-band noise component and of the upper kHz QPO (at least in 4U 1636--53) are hard, whereas the time lags of the lower kHz QPO are soft. As explained by \citet{lee02}, a Compton down-scattering mechanism for the time lags of the lower kHz QPO is also difficult to reconcile with the observed increase in the rms amplitude of the QPO with energy \citep{berger01}, and with the fact that the hard component in the time-averaged spectrum of these systems is interpreted as Compton up-scattering \citep{sunyaev01, payne01}. \citet{lee02} proposed a Compton up-scattering model to explain the soft time lags of the lower kHz QPO in which the temperatures of the Comptonising corona and the source of soft photons oscillate coherently at the QPO frequency \citep[see also][]{lee01}. This could happen, for instance, if the oscillating signal in the soft and hard components are coupled and energy is exchanged between the two. The model of \citet{lee02} is able to reproduce both the energy dependence of the time lags and the rms spectrum of the lower kHz QPO in 4U 1608--52, yielding sizes of the corona of $\sim 4 - 25$ km. Interestingly, since in this model the soft lags are due to the temperature of the corona and that of the source of soft photons oscillating coherently at the same frequency, one would expect that if the time lags of one of the kHz QPOs are soft, the time lags of the other kHz QPO would not (because the signal of the other kHz QPO has a different frequency, and the temperature of the hard and the soft components cannot oscillate coherently at two different frequencies), similar to what we observed. Under this scenario, the signal of the upper kHz QPO would be up-scattered in the hard component, producing the usual hard lags, as we observed. Also, the fact that time lags of the lower kHz QPO are the ones that are soft implies that the exchange of energy between the corona and the source of soft photons is more strongly linked to the oscillations of the lower than those of the upper kHz QPO. We note that in the case of 1H0707--495 \citep{zoghbi01}, while the high-frequency (above $\sim 5 \times 10^{-4}$ Hz) time lags are soft, the low-frequency broad-band time lags are hard, similar to those in galactic black-hole and neutron-star binaries, and hence the same mechanism may be operating in this case.

%
%

The intrinsic coherence of the lower kHz QPO in 4U 1636--53 is consistent with being 0 when the QPO frequency is at $\sim 600$ Hz and increases more or less linearly to 1 as the QPO frequency increases up to 800 Hz (see upper right panel of Figure \ref{fig5}). The intrinsic coherence then remains more or less constant at 1 as the QPO frequency increases further, up to 900 Hz, and then drops abruptly to 0.5 when the QPO frequency is at $\sim 920$ Hz. The intrinsic coherence of the lower kHz QPO in 4U 1608--52 follows a similar trend, albeit being slightly lower than the one in 4U 1636--53, up to $\sim 790$ Hz, where it reaches a value of $\sim 0.6$ (see the same panel of Figure \ref{fig5}). We do not have data at frequencies higher than this for the lower kHz QPO in 4U 1608--52, and therefore we cannot test whether the intrinsic coherence also drops in this source at high QPO frequencies. The behaviour of the intrinsic coherence with QPO frequency in these two sources is remarkably similar to the one of the quality factor, $Q$, of the lower kHz QPO with QPO frequency \citep{barret04}. The quality factor is defined as the ratio of the centroid frequency to the FWHM of the QPO; this quantity is sometimes also called coherence of the QPO, although it is different from the intrinsic Fourier coherence discussed up to here. To avoid confusion, we will call it here either the quality factor or $Q$, and we reserve the term coherence to refer to the intrinsic Fourier coherence of the signal. \citet[][see also \citealt{barret03}]{barret04} proposed that the drop of the quality factor at high QPO frequencies in the lower kHz QPO of 4U 1636--53, 4U 1608--52, and four other sources, reflects the fact that the inner edge of the accretion disc reaches the radius of the ISCO. \citet[][see also \citealt{sanna01}]{mendez04} argued that the drop of $Q$ at high QPO frequencies could be due (at least in part) to changes in the accretion flow around the neutron star that affects the modulation mechanism that amplifies the signal of the QPO at different energies. The striking similarity between the two curves, on one side the intrinsic coherence of the lower kHz QPO vs. QPO frequency (upper right panel of Figure \ref{fig5}) and on the other side the quality factor of the lower kHz QPO vs. QPO frequency \citep[see e.g. Figure 4 of][]{barret05}, provides a strong indication that the two phenomena are related. We reiterate that, in  principle, the two quantities are independent; the quality factor measures the lifetime of the phenomenon that produces the QPO, whereas the intrinsic Fourier coherence measures the degree of correlation between the QPO signal in two different energy bands. While these two quantities can be correlated (as it appears to be the case for these two sources), there is no guarantee that this should always be the case.

An intrinsic coherence of one is equivalent to having two light curves that are related by a linear transform. On the contrary, if the two light curves are related by a non-linear transfer function, power from a given frequency in one light curve is distributed among harmonics of that frequency, leading to a loss of coherence \citep{bendat01, vaughan01}. The loss of intrinsic coherence both at low and high QPO frequencies in 4U 1636--53 (and at low frequencies in 4U 1608--52) entails that for the lower kHz QPO, the $4-12$ keV and $12-20$ keV light curves are related by a non-linear transfer function when the QPO frequency is either low or high. An example of such a signal would be that of two light curves produced by local temperature fluctuations such that the average temperature of the emitting material is much smaller than the energy of the emitted photons. In that case, the two light curves are in the tail of the Wien's spectrum, and they are related to each other via a non-linear transfer function \citep[see][for details]{vaughan01}. Another example of two signals yielding low intrinsic coherence would be that of multiple flaring regions when more than one of those regions contribute to both light curves, even if the individual regions produce perfectly coherent variability \citep{vaughan01}. 

%
%

\cite{zoghbi01} found that the intrinsic coherence between the soft and hard light curves, $0.3 - 1.0$ keV and $1.0 - 4.0$ keV, respectively, in 1H0707--495 are high, $\simmore 0.75$, up to a frequency of $\sim 5 \times 10^{-3}$ Hz. Above that frequency the coherence drops because, according to their simulations, the Poisson noise starts to dominate the variability. It is unclear whether reflection off the accretion disc will always produce a high intrinsic coherence, or how the value of the intrinsic coherence depends on properties of the accretion disc, e.g. the size of the inner radius of the disc. On the other hand, in the model of \citet{lee02} the temperatures of the corona or boundary layer and the accretion disc oscillate coherently, therefore the soft and hard light curves are related by a linear transfer function and the intrinsic coherence is high. The size of the Comptonising medium implied by the fits by \cite{lee02} to the data of 4U 1608--52 suggests that a large region of that medium is involved in the nearly coherent variability, hinting at a global mode in the corona or the boundary layer. If this is correct, the drop of the intrinsic coherence at low and high QPO frequencies would be due to the interplay between the excitation and damping mechanisms that produce this mode. 

On the other hand, the intrinsic coherence of the upper kHz QPO, both in 4U 1608--52 and 4U 1636--53, are consistent with 0 between $500$ Hz and and $1200$ Hz, except for a $\sim 200$ Hz interval around 800 Hz in 4U 1636--53 where the intrinsic coherence increases slightly to 0.13. (We do not have data for the upper kHz QPO of 4U 1608--52 in that frequency range.) In the model of \citet{lee02} this could be interpreted as the source of soft photons and the Comptonising medium not oscillating coherently, which could mean that at those frequencies the source of soft photons and the corona are not (or only weakly) coupled. This could be either because the efficiency of the energy exchange between the source of soft photon on one side and the corona on the other is much higher for producing the lower than the upper kHz QPO, or because the global modes in the system would be more easily excited within a limited frequency range, around 800 Hz. It is interesting that the intrinsic coherence of the upper kHz QPO in 4U 1636--53 appears to increase when the frequency of upper kHz QPO is between $\sim 700$ Hz and $\sim 900$ Hz, peaking at a QPO frequency of $\sim 800$ Hz, the same frequency range over which the intrinsic coherence of the lower kHz QPO in this source increases and peaks. The factor $\sim 10$ difference between the intrinsic coherence of the lower and the upper kHz QPO at $\sim 800$ Hz suggests that the global modes are less efficiently excited, or more efficiently damped, in one case than in the other.


In the case of the lower kHz QPO the intrinsic coherence is relatively high at energies below $\sim 12$ keV where it begins to decline, and then it finally reaches 0 at about 18 keV. \cite{sanna03} analysed six XMM-Newton plus {\em RXTE} spectra of 4U 1636--53 across different spectral states of the source. They fitted the spectra using a model that accounts for the thermal emission form the accretion disc and the neutron-star surface or boundary layer, the Comptonised emission from a corona, plus reflection off the accretion disc. \citet{sanna03} found that in all six observations the Comptonised emission from the corona starts to dominate the spectrum above $\sim 10 -15$ keV. Below this energy, and down to $1-3$ keV, both the neutron-star surface or boundary layer and the corona are more or less equally strong. In the context of the model of \citet{lee02}, our results suggest that the intrinsic coherence is strong over the energy range in which the boundary layer and the Comptonising corona are more or less equally strong, and drops as soon as one of these two components no longer contributes to the total emission.

Finally, in the light of the above discussion, one can speculate whether the same mechanism that produces the soft time lags and drives the loss of the intrinsic coherence of the lower kHz QPO would also affect the quality factor, $Q$, of the QPO. As the temperatures of the source of soft photons and the corona start to oscillate less and less coherently, the feed-back between these two components becomes less and less efficient, which we propose would lead to a decrease of the quality factor and the strength of the QPO, similar to what we observe \citep{barret04, mendez04, sanna01}. The proposal by \cite{barret04} for the drop of $Q$ at high frequencies of the lower kHz QPO involves the inner edge of the accretion disc reaching the ISCO. The drop of $Q$ at low frequencies of the lower kHz QPO is not explained in their model. The mechanism that we propose here would work both at low and high kHz QPO frequencies, and would provide a single explanation to the observed drop of the quality factor (and rms amplitude) of the lower kHz QPOs at both ends of the frequency range. This idea would require further modelling of all the effects involved, which is beyond the scope of this paper.




\section*{Acknowledgments}

The author wish to thank Beike Hiemstra, Diego Altamirano and Guobao Zhang for their help and useful discussions. MGBA and JEH acknowledge the financial support from CAPES. MGBA is grateful to the Kapteyn Astronomical Institute for their hospitality.

\label{lastpage}

\end{document}